\documentclass[aps,preprint,floats,superscriptaddress]{revtex4}

\usepackage{amsmath}
\usepackage{graphicx}
\begin{document}

\title{Transport anomalies in a simplified model for a heavy electron quantum
critical point}

\author{P.~Coleman}

\affiliation{$^{2}$ Center for Materials Theory,
Rutgers University, Piscataway, NJ 08854-8019, U.S.A. }

\author{J.~B.~Marston}

\affiliation{Department of Physics, Brown University, Providence,
Rhode Island 02912-1843, U.S.A.}

\author{A.~J.~Schofield}

\affiliation{School of Physics and Astronomy, University of
Birmingham, Birmingham, B15~2TT, United Kingdom.}

\begin{abstract}
We discuss the transport anomalies associated with the development of
heavy electrons out of a neutral spin fluid using the large-$N$
treatment of the Kondo-Heisenberg lattice model.  At the phase
transition in this model the spin excitations suddenly acquire charge.
The Higgs process by which this takes place causes the constraint
gauge field to loosely ``lock'' together with the external,
electromagnetic gauge field. From this perspective, the heavy fermion
phase is a Meissner phase in which the field representing the
difference between the electromagnetic and constraint gauge field, is
excluded from the material.  We show that at the transition into the
heavy fermion phase, both the linear and the Hall conductivity jump
together.  However, the Drude weight of the heavy electron fluid does
not jump at the quantum critical point, but instead grows linearly
with the distance from the quantum critical point, forming a kind of
``gossamer'' Fermi-liquid.
\end{abstract}
\pacs{}

%
\maketitle
%
\section{Introduction}

The past few years has seen a growth of interest in quantum phase
transitions and the associated phenomenon of quantum
criticality~\cite{continentino_1994a,sachdev_1999a,stewart_2001a,vojta_2003a,
hertz_1976a,millis_1993a,si_2000a,coleman_2001c,coleman_2005a}.
Although quantum criticality develops at a zero-temperature phase
transition, the ability of quantum criticality to profoundly modify
the metallic properties of both d- and f-electron materials at a
finite temperature has aroused intense interest.  Heavy electron
materials have played a particularly important role in the study of
quantum criticality, for these systems lie at the brink of
antiferromagnetic instabilities~\cite{stewart_2001a}, and they are readily
tuned to an antiferromagnetic quantum critical point.  In the approach
to a heavy electron quantum critical point, the characteristic energy
scale of both the Fermi liquid and the magnetic excitations appear to
telescope to zero, as indicated by the appearance of $E/T$ scaling in
inelastic neutron scattering, and the development of scaling in
both the specific heat and resistivity, characterized by a single
energy scale which goes to zero at the quantum critical
point~\cite{schroder_2000a,custers_2003a}.

The standard model for the development of magnetism in a Fermi liquid,
which describes the emergence of antiferromagnetism as a quantum spin
density wave, is unable to account for the simultaneous collapse of
both the Fermi and magnetic energy scales at quantum
criticality~\cite{coleman_2001c}. This suggests the need for a
radically new mechanism for the emergence of magnetism in the heavy
electron state. The solution of classical criticality required two
steps: the formulation of a mean-field theory---provided by the
Landau-Ginzburg theory---and then the application of the
renormalization group to the fluctuations around the mean field.  In
the parallel study of quantum criticality, experiments suggest the
need to search for a new class of mean-field theory that describes the
emergence of magnetism in the heavy electron fluid.  Various new
approaches have been explored, including the idea of local quantum
criticality~\cite{si_2000a}, the notion that spin and charge separate
at a quantum critical point~\cite{coleman_2001c} and the idea that
magnetism develops out of the heavy fermion phase via an intermediate
spin-liquid~\cite{senthil_2004b}.

Part of the difficulty in understanding the heavy electron quantum
critical point stems from the confluence of two separate physical
phenomena. If we consider the passage from the ordered antiferromagnet
into the heavy electron paramagnet there are two processes to
consider:
\begin{itemize}
\item [-] the destruction of ordered magnetism and the 
associated divergence of spin fluctuations in both space and time, and
\item [-] the formation of charged composite heavy electrons associated with
the Kondo quenching of the fluctuating spin degrees of freedom.
\end{itemize}
While a complete theory of heavy electron quantum criticality needs to
unify these two phenomena, Senthil {\it et al.}~\cite{senthil_2004b}
have recently suggested examining models in which these separate
phenomena occur in isolation. They take the view that these two
transitions may actually be physically separated by an intermediate
spin liquid.  Even if this is not the case in practice, we can adopt
their approach as a useful exercise to examine what changes in
transport properties are expected to accompany the formation of
charged heavy electrons.

One of the interesting pieces of physics at a QCP is that the spin
degrees of freedom transform from localized, magnetically polarized
objects, into mobile charged fermions. Very little is known about the
transport anomalies that accompany this transformation.  One proposal
is that the Hall conductivity of the fluid will
jump~\cite{coleman_2001a}.  At present, it is however, only possible
to measure the change in the Hall conductivity at a
magnetically-induced quantum phase transition where recent
measurements suggest that the differential Hall conductivity jumps at
a heavy electron quantum critical point~\cite{paschen_2004a}. In this
paper we examine the model for a heavy electron quantum critical point
proposed by Senthil {\it et al.} and show how the d.c. transport
properties show discontinuities at the transition. In particular the
weak-field Hall effect and the Wiedemann-Franz ratio would jump. By
contrast the optical Drude weight will be continuous through the
transition and reflect a ``gossamer'' Fermi-liquid state.

%
%

\section{Large $N$ Kondo Lattice Model}
Our basic starting model is the large-$N$ fermionic treatment of the
Kondo Heisenberg model~\cite{coleman_1989a} with $N$ spin components labeled by 
Greek indices that run from $1$ to $N$:
\[
H = H_{c}+H_{K}+H_{J} \; ,
\]
where 
\begin{eqnarray}
H_{c}&=& \sum_{i j, \sigma }\left( t_{ij}e^{-ie\int_{j}^{i}\vec A\cdot
d\vec{x}}-\mu \right)c^\dagger_{i \sigma }c^{}_{j \sigma } \; ,
\cr
H_{K}&=& -\frac{J}{N}\sum_{\alpha \beta } 
(c^\dagger_{j\alpha} f^{}_{j\alpha})
(f^\dagger_{j\beta } c^{}_{j\beta }) \; ,
\cr 
H_{H}&=& -\frac{J_{H}}{N}\sum_{(i,j)} ( f^\dagger_{i\alpha }f^{}_{j\alpha}) (
f^\dagger_{j\beta }
f^{}_{i\beta }) \; ,
\end{eqnarray}
describe respectively the conduction band ($H_{c}$), the on-site Kondo
coupling between the local moment and the conduction band 
($H_{K}$), and the super-exchange between neighboring spin sites
($H_{H}$).  A vector potential has been introduced into the hopping
matrix element: for a uniform vector potential, $H_{c}$ can be
rewritten as $H_{c}=\sum \epsilon_{\vec{k}- e \vec{A}}\
c^\dagger_{\vec{k}\sigma}c^{}_{\vec{k}\sigma}$, where the dispersion
$\epsilon_{\vec{k}}= \sum_{\vec{R}}t
(\vec{R})e^{-i\vec{k}\cdot\vec{R}}- \mu$ is the kinetic energy of the
conduction electrons.

The physics of this model depends on the ratio of $x=T_{K}/J_{H}$.  In
the Doniach scenario~\cite{doniach_1977a} for the Kondo lattice, when
$x \ll 1$, the spins antiferromagnetically order, and as $x$ is
increased, the antiferromagnetic state undergoes a transition to heavy
electron paramagnet. The detailed physics of this quantum phase
transition is an unsolved problem.  In a fermionic mean-field
approach, valid in the large-$N$ limit, antiferromagnetism at small
$x$ is replaced by a spin-liquid or valence-bond ground-state\cite{coleman_1989a,affleck_1988a,vojta_1999a}.
Although this model does not permit us to examine the destruction of
magnetism by the Kondo effect, it nevertheless affords the
opportunity to examine the change in transport properties which
accompany the formation of the heavy electron paramagnet. This is the
topic of this paper.

When we formulate the Kondo Heisenberg model as a functional integral,
we can decouple the Kondo and Heisenberg  terms 
into a ``slave boson'' and an RVB gauge field, as follows:
$H_K \rightarrow H'_K$ and $H_H \rightarrow H'_H$ where
\begin{eqnarray}
H_{K}'&=&\sum_{j} \left[ \bar
V_{j} (c^\dagger_{j\alpha}f^{}_{j\alpha})
+ V_{j} (f^\dagger _{j\beta}c_{j\beta
})
\right] + \frac{N\bar V_{j}V_{j}}{J} \; ,
\nonumber \\
H_{H}'&=&\sum_{(i,j)} ( \left[\vert \chi_{ij}\vert e^{-i
\int_{j}^{i}\vec{\theta}\cdot d\vec{x}} f^\dagger _{i\alpha }f_{j\alpha}
+ {\rm H. c. } \right] + \frac{N\vert \chi _{ij}\vert ^{2}}{J_{H}}.
\nonumber
\end{eqnarray}
Here the bond variable, $\chi$, in the second term has been written
as the product of an amplitude and phase term. The gauge field $\theta
_{ij}= \int _{j}^{i}\vec{\theta }\cdot d\vec{x}$ has been written in a
form which assumes smooth variations about a uniform
configuration. Fluctuations of this gauge field enforce the constraint
that the spinon current flow between sites is zero.  The mean-field
theory of this model, requires an additional constraint term, as
follows
\[
H=H_{c}+ H_{K}'+ H_{H}'+ \sum _{j}\lambda (f^\dagger _{j\sigma}f_{j\sigma} - Q) \; ,
\]
so that variations in $\lambda$ enforce the constraint $n_{f}=Q$,
where $Q\sim O (N)$ is the number of f-electrons used to represent the
large $N$ spin.

There are two mean-field phases of interest, 
\begin{itemize}

\item A uniform RVB spin liquid phase where $\langle V\rangle =0 $, but $\chi
_{ij}=\chi $ assumes a uniform value.  In this phase the spinons,
represented by $f$ fermions are unconfined, and have a dispersion determined
by $j_{\vec{k}}= \chi \sum_{\vec{R}}
e^{{i\vec{k}\cdot\vec{R}}}+\lambda$.

\item The heavy electron phase, where $\langle V_{j}\rangle =V \ne 0$ and 
$\chi _{ij}= \chi $ are both finite and uniform. 

\end{itemize}

\begin{figure}
\includegraphics[width=\columnwidth]{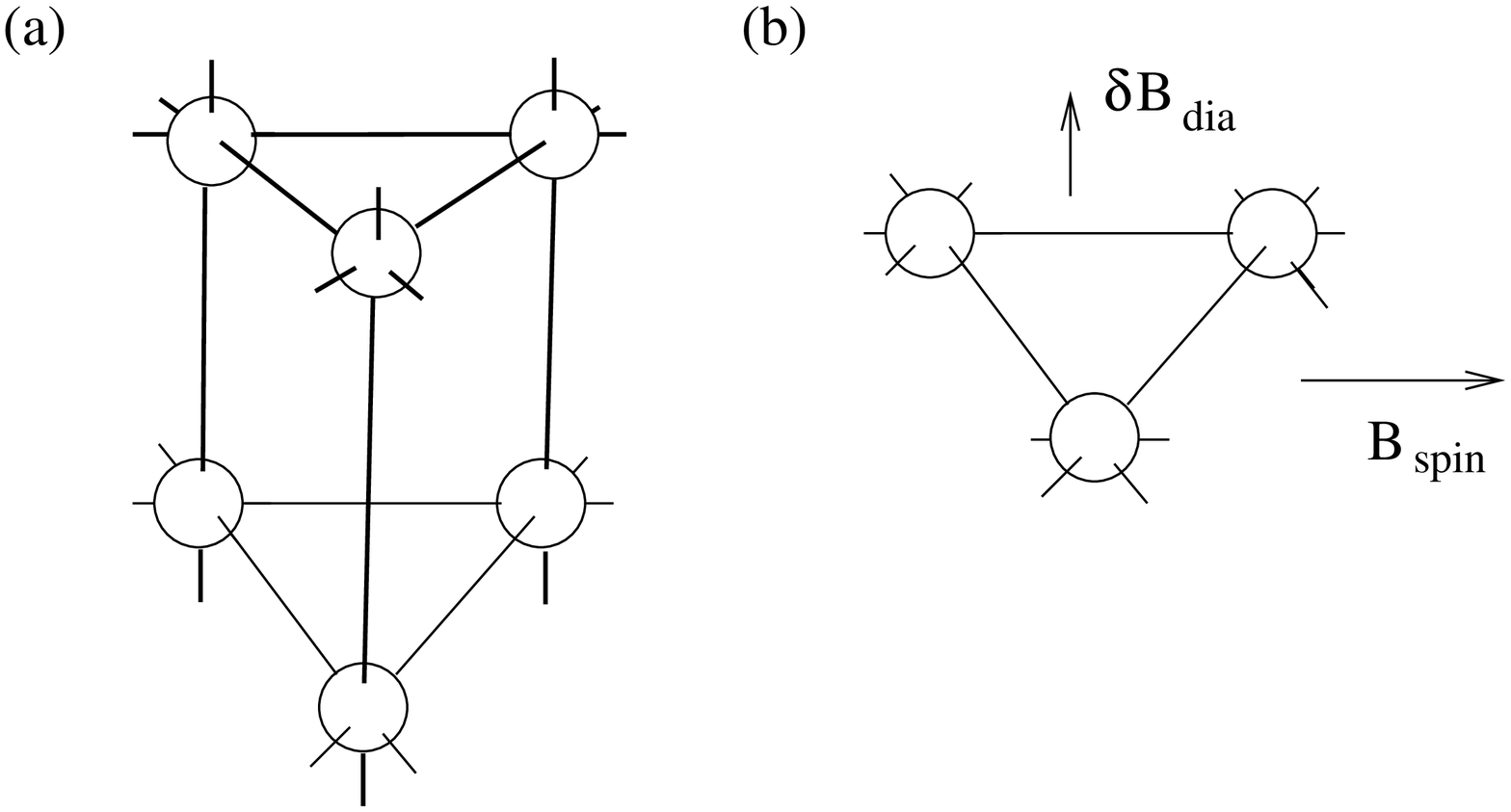}
\caption{\label{fig1}
(a) 3D triangular spin structure for which the uniform RVB phase is
stable in the large $N$ limit. 
(b) The simpler 2D triangular lattice has the conceptual advantage
that spin and diamagnetic parts of the magnetic field can be treated
separately by considering fields parallel and perpendicular to the
plane.}
\end{figure}

In this discussion, our main focus is on the transformation of the
transport properties which occur at the second-order transition
between these two phases.  Strictly speaking, stabilization of the
uniform RVB phase against instabilities to valence bond states, flux
phases or plaquet states requires the presence of additional
frustrating interactions, such as bi-quadratic Heisenberg couplings or
spin-exchange terms around plaquets. Our discussion will assume that
the Heisenberg terms are sufficiently frustrated to permit us to
analyze a second-order between the two uniform phases. As a specific
example, of such a model, we might consider a triangular crystal
structure the type shown in Fig. \ref{fig1}, where frustration in the
layers stabilizes the uniform solution. (To avoid a flux phase, we
would need in addition, to include a ring exchange term around each
rhombohedral plaquet.)  It is conceptually easier in the model
discussion we develop here, to take the two dimensional limit of this
model. This has the conceptual advantage that magnetic field parallel
to the layer couples only to the spin part of the Hamiltonian, so that
we can separate the spin effects of field tuning from the diamagnetic
effects associated with a field perpendicular to the plane.

To discuss the transport, we shall suppose that there is weak site
disorder in the conduction electron fluid, and weak bond- disorder in
the spin liquid to provide an elastic scattering mechanism.

\section{Effective action and the coupling of Gauge Fields}

\begin{figure}
\includegraphics[width=\columnwidth]{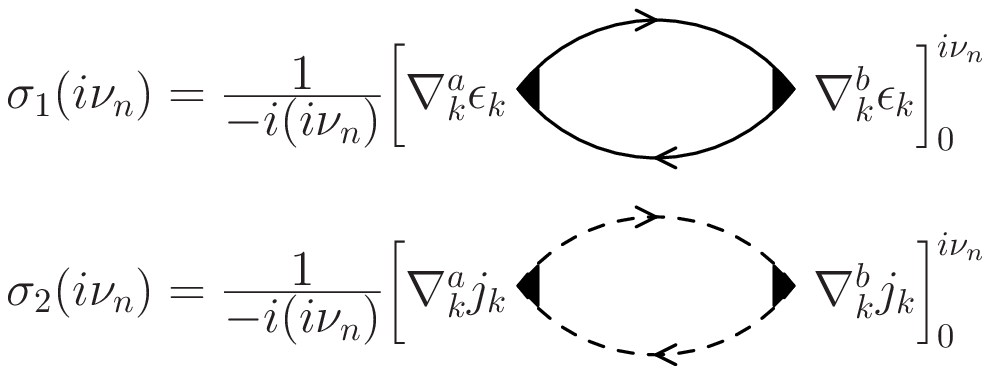} 
\caption{\label{fig2} Diagrams for the conductivities entering into
the effective action. Full propagators refer to conduction electron
propagators $G_{c}^{-1} (i\omega_{n})= i \omega_{n}- \epsilon_{k}+
i{\rm sgn} (\omega_{n}) / (2\tau_{1}) $, and dashed lines refer to
f-electron propagators $G_{f}^{-1} (i\omega_{n})= i \omega_{n}- j_{k}+
i{\rm sgn} (\omega_{n}) / (2\tau_{2}) $, }
\end{figure}

In the spin-liquid phase, where $\langle V\rangle =0$, the
$\theta $ field decouples from the external vector potential. Just as
the physical vector potential couples to currents of the conduction
electrons, the $\theta $ field couples to ``currents'' of the f-spinons.
For slow variations of these fields, the effective action
obtained by integrating out the fermions, will take the form
\begin{equation}
{\cal S}_{o} =  \frac{1}{2}\int \! \! \frac{d\omega}{2\pi } 
\left[-i\omega \sigma _{1}
(\omega) e^{2}\vert A (\omega)\vert ^{2} -i  \omega \sigma _{2}
(\omega) \vert \theta  (\omega)\vert ^{2}  \right] .
\end{equation}
In the relaxation
time approximation the conductivities
\begin{eqnarray}
\sigma _{1} (\omega) &=& \frac{\Omega_{1}^{2}}{\tau _{1}^{-1}- i
\omega} \; ,\\
\sigma _{2} (\omega) &=& \frac{\Omega_{2}^{2}}{\tau _{2}^{-1}- i \omega} \; ,
\end{eqnarray}
describe the Drude response of the conduction and spinon fluid, to
their respective gauge fields.  The diagrams involved in the formal
evaluation of these two quantities are shown in Fig.~\ref{fig2}.

Although the vector potential and the $\theta $ fields couple in an
essentially identical way to their respective fluids, $\vec{\theta }$ is
not an external field like $\vec{ A}$, but fluctuates around a mean
value $\vec{\theta }=0$.  Fluctuations over this field guarantee the
neutrality of the spin-liquid and preclude any flow of charge
associated with the super-exchange couplings.

The transition between the spin-liquid and the heavy electron state
involves the development of a small slave boson amplitude via a
second-order phase transition. This occurs when the free energy, $F$,
satisfies
\begin{eqnarray}
0&=&\frac{\partial^{2}F}{\partial \bar V \partial V }= 
\frac{N}{J}- N\sum _{k} \frac{f(j_{k})-f
(\epsilon_{k})}{\epsilon_{k}-j_{k}} \\
&\sim& \frac{N}{J}- N\rho \ln \frac{D}{J_{H}}
= -N \rho \ln \frac{T_{K}}{J_{H}} \; ,
\end{eqnarray}
where $\rho$ is the density of conduction electrons, and the estimate
of the integral has been made, assuming that the band-width $D$ of the
conduction electrons is much greater than the band-width of the spin
liquid. So, when $T_{K}> (T_{K})_{c}\sim J_{H}$, the spin liquid
becomes unstable with respect to the heavy electron state. 

On the heavy electron side of this transition we have $\langle V
\rangle \neq 0$ and the dispersion of the unconfined spinons and
electrons now become mixed, forming a quasiparticle dispersion of the
form
\[
E_{\vec{k}}^{\pm} = \frac{1}{2} (\epsilon_{\vec{k}}+ j_{\vec{k}})\pm
\sqrt{
\left(\frac{(\epsilon_{\vec{k}}- j_{\vec{k}})}{2}  \right)^{2} + V^{2}
} \; .
\]
In general, the point where $j_{\vec{k}}=\epsilon_{\vec{k}}$ will in
general be far from the Fermi surface, so that for small $V$, the
Fermi surfaces of the heavy electron fluid are essentially identical
to conduction sea plus spin fluid. Nevertheless, the presence of a
small slave boson amplitude leads to a non-trivial coupling between
the physical vector potential $\vec{A}$ and the gauge field
$\vec{\theta }$, given by
\begin{equation}\label{coupling}
{\cal S}_{c} =  a  V^{2}\int \frac{d\omega}{2\pi } 
\vert e \vec{A} (\omega) - \vec{ \theta } (\omega)\vert ^{2}.
\end{equation}
To understand this coupling, consider the physical electromagnetic
gauge transformations in the heavy electron phase. We can always
choose a gauge where $V=\vert V\vert $ is real.  In the heavy electron
phase the hybridization between the f-spinons and conduction electrons given
by $H_{K}'$, means that invariance of the Lagrangian only occurs if
one carries out a single gauge transformation applied to both the f- and
conduction fermions, i.e.  the Lagrangian is only invariant under a
single joint gauge transformation
\begin{equation}
\begin{array}{ll}
c_{j\sigma }\rightarrow c_{j\sigma } e^{i \phi (\vec{R}_{j})}\; , &
\vec{ A} \rightarrow  \vec{A} + \frac{1}{e} \vec{\nabla }\phi \; ,\\
f_{j\sigma }\rightarrow  f_{j\sigma } e^{i \phi (\vec{R}_{j})}\; ,& 
\vec{ \theta } \rightarrow  \vec{\theta } + \vec{\nabla }\phi \; .
\end{array}
\end{equation}
This implies that when we expand the mean-field free energy in
$\vec{A}$ and $\vec{\theta }$, the free energy will be strictly a function of
$\vec{A}-\vec{\theta }$. So, to quadratic order in the fields, 
the full Lagrangian for the gauge field couplings is 
\begin{eqnarray}\label{lg}
{\cal S} &=& \frac{1}{2} \int \frac{d\omega}{2\pi } 
\left[-i\omega \sigma _{1}
(\omega) e^{2}\vert A (\omega)\vert ^{2} -i  \omega \sigma _{2}
(\omega) \vert \theta  (\omega)\vert ^{2}  \right] \nonumber \\
&&+ \frac{1}{2} \int \frac{d\omega}{2\pi}
a V^{2}
\vert e \vec{A} (\omega) - \vec{ \theta } (\omega)\vert ^{2}.
\end{eqnarray}

\begin{figure}
\includegraphics[width=\columnwidth]{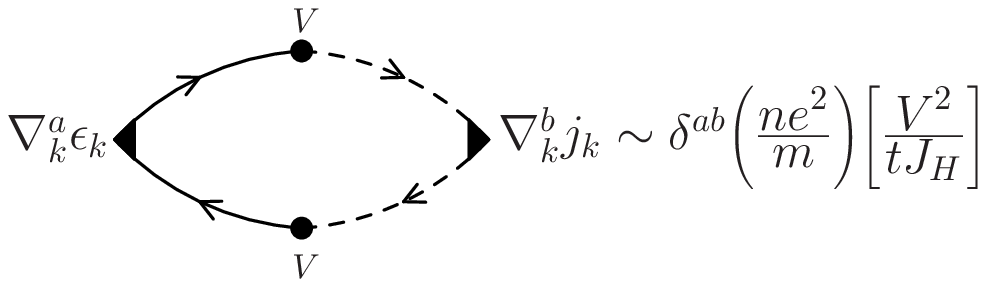}
\caption{\label{fig3}
The hybridization $V$ induces a further response to an applied
electromagnetic field that leads to an additional term to the physical 
conductivity illustrated here.}
\end{figure}

The coupling term between the two fields is given by the diagram shown
in Fig.~\ref{fig3}.  Here $n$ and $m$ are respectively, the number
density and effective mass of the conduction electrons, so that $a
\sim \frac{n}{m t J_{H}}$.  In Eq.~(\ref{lg}) we see that that both gauge
fields, $\vec{A}$ and $\vec{\theta }$, have developed a mass $\propto
V^{2}$ so that, at first sight, both the conduction fluid and the spin
liquid have becomes superfluids.  However, the linear coupling term $
a e \vec{ A}\cdot \vec{ \theta } $ between them ensures that, at long
times, $\theta $ adapts to the value $\theta= e \vec{A} $ which minimizes
the Free energy. It is this term that gives charge to the f-electrons
so that the supercurrents of conduction and f-electrons mutually
screen one-another.  From the Gaussian coefficient of $\theta $, we see
that the propagator associated with the $\theta $ field is
\begin{equation}
D_{\theta } (\omega )^{-1 } = - i \omega \sigma _{2}  + a V^{2} \; ,
\end{equation}
indicating that the characteristic response rate of this field is
given by
\begin{equation}
\tau _{*}^{-1} = a V^{2} /\sigma_{2} \; .
\end{equation}
This sets the rate at which the supercurrents in the f- and conduction
fluid adapt  to screen one another. On time-scales longer than $\tau
_{*}$, the fluid is  a heavy paramagnet. 

One of the interesting consequences of the spin fluid acquiring a
charge, is that the d.c. conductivity of the fluid must jump at the
transition from spin to heavy electron liquid\cite{houghton_2002a}.  When the spin liquid
is neutral (the $|V|=0$ phase), the d.c. conductivity is simply given
by the conduction electron component
\begin{equation}
\sigma ^{-}= e^{2}\sigma _{1} + 0. \sigma _{2} = e^{2 }\sigma _{1} \; .
\end{equation}
However, once the spins acquire a charge in the heavy electron phase
then the spinon fluid contributes directly to the
D.C. conductivity
\begin{equation}
\sigma ^{+}= e^{2} (\sigma _{1} + \sigma _{2}) \; .
\end{equation}

By contrast, we expect the thermal conductivities to be unchanged by
the transition, because thermal currents of the spin liquid do not
couple to the gauge fields.  If $\kappa_{1}$ and $\kappa_{2}$ are the
thermal conductivities of the conduction and spin fluids just before
the transition, then the thermal conductivity of the heavy electron
fluid will be given by $\kappa_{HF}= \kappa_{s}+\kappa_{e}$ on either
side of the transition.  From this discussion, we see that at
transition from heavy electron fluid to spin liquid the
Wiedemann-Franz ratio will jump from the standard value
\begin{equation}
\frac{\kappa_{+}}{\sigma_{+} T}=W= 
\frac{\pi^{2}}{3}\left(\frac{k_{B}}{e} \right)^{2} \; ,
\end{equation}
in the heavy electron fluid, to the larger value
\begin{equation}
\frac{\kappa_{1}+\kappa_{2}}{e^{2}\sigma _{1}T} = W
\left[1+\frac{\sigma _{2}}{\sigma _{1}} \right] \; ,
\end{equation}
when the spin  liquid becomes neutral.

We can calculate the jump in the d.c. conductivity directly by
integrating out the $\vec{\theta} $ field from Eq.~(\ref{lg}). This
gives
\begin{eqnarray}\label{lg*}
{\cal S}^{*} &=& \frac{e^{2}}{2} \int \frac{d\omega}{2\pi } 
\left\{[ -i\omega \sigma _{1} + a V^{2}- ( a V^{2})^{2}D_{\theta }
(\omega) 
] \vert A (\omega)\vert ^{2}
 \right\}\cr
& =& \frac{1}{2} \int \frac{d\omega}{2\pi } 
\left\{[ -i\omega\sigma _{+} (\omega)
] \vert A (\omega)\vert ^{2}
 \right\} \; ,
\end{eqnarray}
where 
\begin{equation}\label{lg**}
\sigma _{+} = e^{2}\left[\sigma _{1} (\omega)+ \frac{aV^{2}}{
\tau_{*}^{-1}-i\omega
} \right]  \; ,
\end{equation}
is the renormalized conductivity. In this expression we have have
omitted terms of order $aV^{2}/\Omega_{2}^{2}$ in the denominator of
the second term. From this, we see that the optical conductivity
acquires a new Drude peak at low energies, of width $\tau _{*}^{-1}=
aV^{2}/\sigma _{2}$, weight $aV^{2}$. The zero frequency limit of this
expression is indeed $e^{2} (\sigma _{1}+\sigma _{2})
$\cite{sumrule}.

\section{Jump in the Hall conductivity}

To discuss the Hall conductivity, we need to consider the response of
the system to crossed magnetic and electric fields which implies that
we must consider the cubic interaction terms between the gauge fields.
The Hall response requires us to 
to consider configurations
of the vector potential containing a spatially uniform electric
component, $A^E(t)$ and a time-independent magnetic component
$A^B(\vec x)$, such that $\vec E = - \partial A^E/\partial t$ and
$\vec B = \vec \nabla \times \vec A$,
\begin{equation}
\vec{A} (\vec{x},t) = \vec{A}^{P} (\vec{x},t) +\vec{A}^{B}
(\vec{x})+\vec{A}^{E} (t) \; ,
\end{equation}
where $\vec{A}^{P}$ is the residual ``probe'' part of the vector
potential that is neither constant in space or time.
In Fourier space,
\begin{equation}\label{mixfields}
\vec A(\vec q, \omega) = \vec A^P(\vec q, \omega) + \vec A^E(\omega) (2 \pi)^3 
\delta^{(3)}(\vec q) + \vec A^B(\vec q) 2 \pi \delta(\omega)
\end{equation}
where $\vec A^P$ is the field used to probe the current. We also need
to consider the analogous gauge fields that constrain the circulating
currents in the spin liquid.  The crossed gauge fields will now
introduce a cubic term
\begin{eqnarray}\label{cubic}
{\cal S}^{3} &=& \int \frac{d^3 q}{(2 \pi)^3}
\left[
A^P_a(-q)A^B_b(\vec q)A^E_c(\omega) Q_{abc}(q) \right. \nonumber \\ 
&& + \left.
\theta ^P_a(-q)\theta ^B_b(\vec q)\theta ^E_c(\omega) P_{abc}(q) 
\right] \; ,
\end{eqnarray} 
where $q\equiv (\vec{ q},\omega )$. 
The coefficients $Q_{abc}$ and $P_{abc}$ are related to the
three-point current fluctuations of the conduction and spin fluids
respectively, 
\begin{eqnarray}\label{currentflucs}
Q_{abc} (q) &=& \langle J_{a} (-q)J_{b} (\vec{q})J_{c} (\omega)\rangle
_{q}\nonumber 
\\ 
&&- \langle J_{a} (-q)J_{b} (\vec{q})J_{c} 
(\omega)\rangle _{q=0} \; ,
\\
P_{abc} (q) &=& \langle {\cal J}_{a} (-q){\cal J}_{b}
(\vec{q}){\cal J}_{c} (\omega)\rangle _{q}\nonumber \\
&&- \langle {\cal J}_{a}
(-q){\cal J}_{b} (\vec{q}){\cal J}_{c} (\omega)\rangle _{q=0}. 
\end{eqnarray}
Here the currents, 
\begin{eqnarray}
\vec{J}_{a} (\vec{q}) &=&e\sum_{\vec{k}
\sigma }\vec{\nabla
}\epsilon_{\vec{k}}c^\dagger _{\vec{k}-\vec{q}/2\sigma }c_{\vec{k}+\vec{q}/2
\sigma } \; , \\ 
\vec{\cal J}_{a} (\vec{q}) &=& \sum_{\vec{k}
\sigma }\vec{\nabla
}j_{\vec{k}}f^\dagger _{\vec{k}-\vec{q}/2\sigma }f_{\vec{k}+\vec{q}/2\sigma }
\; , 
\end{eqnarray}
are the ``paramagnetic'' current fluctuations of the conduction
electrons and spin liquid, respectively.  In an isotropic conduction
fluid, where the Hall conductance $\sigma _{xy}= \alpha B$ gauge
invariance requires that
\begin{eqnarray}
\vec{J}_{a}&=& \alpha (\vec{E}\times \vec{B})=  
\alpha \biggl[(- i \omega
A^{E})\times (i \vec{q}\times \vec{A}^{B})\biggr]\cr
&=& - \alpha \omega \biggl(
q_a \delta_{bc} - q_b \delta_{ac}\biggr) A^{E}_{b}A^{B}_{c} \nonumber \\
&\equiv& -
Q_{abc}A^{E}_{b}A^{B}_{c}
\end{eqnarray}
so that we can identify the coefficient of the Hall conductance
\begin{eqnarray}
Q_{abc}(\vec q,\omega) = 
\alpha
\omega \biggl(
q_a \delta_{bc} - q_b \delta_{ac}\biggr) \; .
\end{eqnarray}
In the spin liquid phase, the absence of any coupling between the two
gauge fields guarantees that $\sigma _{xy}= \alpha B$ is the Hall
response of the combined system.  Similarly, it can be shown that
\begin{eqnarray}
P_{abc}(\vec q,\omega) = 
\gamma
\omega \biggl(
q_a \delta_{bc} - q_b \delta_{ac}\biggr) \; ,
\end{eqnarray}
describes the analogous quantity for the spin liquid. 
In a simple relaxation time approximation, the above coefficients can
be related to Fermi surface integrals of the quasiparticle mean-free
path around the two dimensional Fermi surface, given by~\cite{ong_1991a}
\begin{eqnarray}
\alpha &=& \frac{e^{3} (N/2))}{2\pi^{2 }\hbar }\oint d\vec{ l}_{\vec{k}}
\times \vec{l}_{\vec{k}} \; , \nonumber \\
\gamma &=& \frac{ (N/2))}{2\pi^{2 }\hbar }\oint d\vec{ {\cal
L}}_{\vec{k}}\times \vec{{\cal  L}}_{\vec{k}} \; ,
\end{eqnarray}
where $\vec{l}_{\vec{k}}=\vec{v}_{\vec{k}}\tau _{\vec{k}}$ and
$\vec{\cal L}_{\vec{k}}= \vec{v}_{f\vec{k}}\tau _{f\vec{k}}$ are the
mean-free paths of the conduction and spin quasiparticles
respectively.  Now suppose we cross through the quantum phase
transition into the heavy electron paramagnet.  In this case, close to
the transition, when integrate out the fluctuations over $\theta $, we
must replace $\theta $ by its expectation value
\begin{equation}
\vec{\theta } (q) \rightarrow  (aV^{2}) e D (q)\vec{A} (q) \; .
\end{equation}
The renormalized cubic term is then given by
\begin{eqnarray}\label{cubicstar}
{\cal S}^{3*} = \int \frac{d^3 q}{(2 \pi)^3}
\left[
A^P_a(-q)A^B_b(\vec q)A^E_c(\omega) Q^{*}_{abc}(q) 
\right] \; ,
\end{eqnarray} 
where
\begin{equation}
Q^{*}_{abc}(q) = 
Q_{abc}(q) +
 (e aV^{2})^{3} D (-q)D (\vec{q},0) D (0,\omega) P_{abc} (q) .
\end{equation}
But in the long-wavelength, low frequency  limit $D (q)\rightarrow 1/
(aV^{2})$, so that in this limit $\theta (q)\rightarrow eA (q)$ and 
\begin{equation}
Q^{*}_{abc}(q) \rightarrow 
Q_{abc}(q) +
 e^{3}P_{abc} (q) \; .
\end{equation}
In other words, we expect the  Hall
conductivity of the heavy electron fluid to be
\begin{equation}
\sigma ^{*}_{xy}= (\alpha +e^{3} \gamma)B , 
\end{equation}
so that the Hall conductivity $\sigma _{xy}$ jumps by an amount
$\Delta \sigma _{xy}= e^{3}\gamma B$.  This jump in the Hall response
can be traced back to the fact that the previously neutral
heavy-electron currents now become charged. At a zero field quantum
critical point, the linear Hall current induced by a tiny, but fixed
field will actually jump.

\section{Discussion}

In this paper, we have studied a simplified model for a heavy electron
quantum phase transition, involving a transition from a metal plus
decoupled spin liquid, to a heavy electron fluid in which the
fermionic spin excitations develop charge. Although our model is
grossly over-simplified, it does illustrate how the development of
charged heavy electrons out of the spin fluid is expected to affect
the transport.

There are a number of interesting questions and observations that
emerge from our model treatment. 
\begin{itemize}

\item The primary conclusion from our study is that the
d.c. electrical conductivities (both longitudinal and Hall components) are
expected to jump discontinuously at a quantum critical point between a
heavy fermion metal and a spin liquid. The key ingredient in this
transition is the sudden appearance of resonant levels at each site in
the lattice which scatter the conduction electrons.

\item Our simple model assumed a separation between the spin and
orbital parts of the magnetic field.  Real heavy electron systems
exhibit strong spin orbit coupling, so that the idealized separation
between the spin and orbital parts of the magnetic field can not in
general, be made. Nevertheless, the discontinuities we have found
should be a general feature of any transition where the f-spin
excitations suddenly develop a charge.

\item If we contrast our results with those anticipated in a spin
density wave transformation of the Fermi surface, we note that here
the magnetic order will couple linearly to the conductivity and to the
Hall conductivity, so that gradual evolutions in the Hall constant at
the quantum phase transition are expected in the limit of weak
magnetic fields\cite{fenton}.  Thus a jump in the Hall conductivity
is a sign that the order parameter is intimately connected with the
formation of new, coherent propagating charged quasiparticles.

\end{itemize}

\acknowledgments The authors would like to thank the hospitality of
the Max Planck Institute for Complex Systems, Dresden, the Aspen
Center for Physics and the Kavli Institute for Theoretical Physics,
Santa Barbara, where part of the work in this article was carried out.
This work was supported by NSF grants DMR 0312495 (PC) and DMR 0213818
(JBM), The Royal
Society and Leverhulme Trust (AJS).  This research was also
supported in part by the National Science Foundation under Grant
No. PHY99-0794.  We would like to thank Dr. Silke Paschen for
discussing her measurements on heavy electron materials prior to
publication.  


\begin{thebibliography}{21}
\bibitem[{\citenamefont{} ()}]{sumrule}
\bibinfo{note}{At
first sight, our result appears to give a slight violation of the
optical sum rule, since now $\int \frac{d\omega}{2\pi} \sigma _{+}
(\omega) = e^{2} \Omega_{1}^{2} + e^{2} (aV^{2})$. However, this can
be traced back to the fact that we have not allowed for the slight
renormalization of the dissipative part of Eq.~(\ref{coupling} )
$\Omega_{1}^{2}\rightarrow \Omega_{1}^{2}- aV^{2}$ that is required in
order to keep the high frequency coefficient of $\vert A (\omega)\vert
^{2}$ unchanged in (\ref{lg}). This small change means that the weight
for the low frequency Drude peak is transfered from high frequencies
of order $\omega\sim (\tau _{1})^{-1}$.
}
\end{thebibliography}
%
%

\end{document}